%% file: ms.tex
\documentclass[letterpaper, 10 pt, conference]{ieeeconf} 
\IEEEoverridecommandlockouts   % This command is only needed if you want to use the \thanks command

\usepackage{cite}
\usepackage{amsmath,amssymb,amsfonts}
\usepackage[normalem]{ulem} % for \sout
\usepackage{xcolor}
\usepackage{booktabs}
\usepackage{subcaption}
\usepackage{etoolbox}
\newbool{long}
\setbool{long}{false}

\newcommand{\inputnewtext}[1]{\ifbool{long}{\textcolor{blue}{#1}}{#1}}
\newcommand{\deleteoldtext}[1]{\ifbool{long}{\textcolor{red}{#1}}{}}

\def\BibTeX{{\rm B\kern-.05em{\sc i\kern-.025em b}\kern-.08em T\kern-.1667em\lower.7ex\hbox{E}\kern-.125emX}}
\usepackage{amsthm}% Formats (bold, no indent) theorems, comments and stuff
\newtheoremstyle{dotless}{}{}{\itshape}{}{\bfseries}{}{ }{}

\let\labelindent\relax % Clash with enumitem
\usepackage{enumitem}

\usepackage{graphicx}
\usepackage{pgfplots}
\pgfplotsset{compat=newest}

\theoremstyle{dotless}
\newtheorem{dummy}{}

\newtheorem{theorem}[dummy]{Theorem}

\newtheorem{lemma}[dummy]{Lemma}
\newtheorem{procedure}[dummy]{Algorithm}

\newtheorem{definition}[dummy]{Definition}
\newtheorem{remark}[dummy]{Remark}

\newcommand{\abs}[1]{\left\lvert #1 \right\rvert}

\newcommand{\T}{T}
\newcommand{\diag}{\textnormal{diag}}

\renewcommand{\vec}[1]{\textnormal{vec}(#1)}
\newcommand{\uhat}{\hat{u}(k \vert k-1)}

\newcommand{\gradphij}{\frac{d}{d\phi_j}}

\input{my_colors.tex}

\newcommand{\drawlinelegend}[1]{\raisebox{.5ex}{\tikz{\draw[#1, line width=0.4mm] (0,0) -- +(1em, 0);}}}
\newcommand{\drawrectanglelegend}[2]{\raisebox{.0ex}{\tikz{\filldraw[color=#1!#2, fill=#1!#2] (0,0) rectangle (2ex, 1ex);}}}

\ifbool{false}{}{
\renewcommand{\baselinestretch}{0.97} 
}
\begin{document}

\bstctlcite{IEEEexample:BSTcontrol} % Fix ---- for author names in bibliography

% \title{\LARGE \bf Learning the Scheduling Coefficients in Feedforward Control: \\ a Neural Network Approach
% }

%\title{\LARGE \bf Direct Learning of Linear Parameter Varying Feedforward Control for Nonlinear systems}

%\title{\LARGE \bf Parameter-Varying Feedforward Control: \\ A Neural-Network Scheduling Approach}

% \title{\LARGE \bf Direct Learning of Parameter Varying Feedforward Control: \\ A Neural-Network Scheduling Approach}

\title{\LARGE \bf Direct Learning for Parameter-Varying Feedforward Control: \\ A Neural-Network Approach}

\author{Johan Kon$^1$, Jeroen van de Wijdeven$^2$, Dennis Bruijnen$^3$, Roland T\'{o}th$^{4}$, Marcel Heertjes$^{1,2}$, Tom Oomen$^{1,5}$ % <-this % stops a space
\thanks{This work is supported by Topconsortia voor Kennis en Innovatie (TKI), and ASML and Philips Engineering Solutions. $^1$: Control Systems Technology Group, Departement of Mechanical Engineering, Eindhoven University of Technology, Eindhoven, The Netherlands, e-mail: j.j.kon@tue.nl. $^2$: ASML, Veldhoven, The Netherlands. $^3$: Philips Engineering Solutions, Eindhoven, The Netherlands. $^4$: Control Systems Group, Electrical Engineering, Eindhoven University of Technology, The Netherlands, and the Institute for Computer Science and Control, Budapest, Hungary. $^5$: Delft University of Technology, Delft, The Netherlands. }% <-this % stops a space
}

% m.f.heertjes@tue.nl, t.a.e.oomen@tue.nl.
% e-mail: dennis.bruijnen@philips.com 
\maketitle

\input{./Sections/Abstract.tex}

\input{./Sections/Introduction.tex}

\input{./Sections/Problem_formulation.tex}

\input{./Sections/Local_optimization.tex}

\input{./Sections/Pseudolinear_regression.tex}

\input{./Sections/Simulation_example.tex}

\input{./Sections/Conclusion.tex}

\bibliographystyle{IEEEtran}
\bibliography{IEEEabrv,BSTcontrol.bib,library.bib}

\end{document}

%% file: my_colors.tex
\definecolor{red}{rgb}{1,0,0}
\definecolor{blue}{rgb}{0,0,1}
\definecolor{green}{rgb}{0,1,0}
\definecolor{yellow}{rgb}{1,1,0}
\definecolor{orange}{rgb}{1,0.647,0}
\definecolor{gold}{rgb}{1,0.843,0}
\definecolor{purple}{rgb}{0.627,0.125,0.941}
\definecolor{gray}{rgb}{0.745,0.745,0.745}
\definecolor{brown}{rgb}{0.647,0.165,0.165}
\definecolor{navy}{rgb}{0,0,0.502}
\definecolor{pink}{rgb}{1,0.753,0.796}
\definecolor{seagreen}{rgb}{0.18,0.545,0.341}
\definecolor{turquoise}{rgb}{0.251,0.878,0.816}
\definecolor{violet}{rgb}{0.933,0.51,0.933}
\definecolor{darkblue}{rgb}{0,0,0.545}
\definecolor{darkcyan}{rgb}{0,0.545,0.545}
\definecolor{darkgray}{rgb}{0.663,0.663,0.663}
\definecolor{darkgreen}{rgb}{0,0.392,0}
\definecolor{darkmagenta}{rgb}{0.545,0,0.545}
\definecolor{darkorange}{rgb}{1,0.549,0}
\definecolor{darkred}{rgb}{0.545,0,0}
\definecolor{lightblue}{rgb}{0.678,0.847,0.902}
\definecolor{lightcyan}{rgb}{0.878,1,1}
\definecolor{lightgray}{rgb}{0.827,0.827,0.827}
\definecolor{lightgreen}{rgb}{0.565,0.933,0.565}
\definecolor{lightyellow}{rgb}{1,1,0.878}
\definecolor{black}{rgb}{0,0,0}
\definecolor{white}{rgb}{1,1,1}
\definecolor{mblue}{rgb}{0, 0.4470, 0.7410}
\definecolor{mred}{rgb}{0.85, 0.325, 0.098}
\definecolor{morange}{rgb}{0.929, 0.694, 0.125}
\definecolor{mpurple}{rgb}{0.494, 0.184, 0.556}
\definecolor{mgreen}{rgb}{0.466, 0.674, 0.188}
\definecolor{mcyan}{rgb}{0.301, 0.745, 0.933}
\definecolor{mbrown}{rgb}{0.635, 0.078, 0.184}

%% file: Sections/Abstract.tex
\begin{abstract}
The performance of a feedforward controller is primarily determined by the extent to which it can capture the relevant dynamics of a system. The aim of this paper is to develop an input-output linear parameter-varying (LPV) feedforward parameterization and a corresponding data-driven estimation method in which the dependency of the coefficients on the scheduling signal are learned by a neural network. The use of a neural network enables the parameterization to compensate a wide class of constant relative degree LPV systems. Efficient optimization of the neural-network-based controller is achieved through a Levenberg-Marquardt approach with analytic gradients and a pseudolinear approach generalizing Sanathanan-Koerner to the LPV case. The performance of the developed feedforward learning method is validated in a simulation study of an LPV system showing excellent performance.
\end{abstract}

%% file: Sections/Introduction.tex
\section{Introduction}
\label{sec:introduction}
Feedforward control can substantially improve the performance of dynamic systems performing tracking tasks \cite{Clayton2009}, e.g., the position accuracy in motion systems. To obtain flexibility with respect to varying tasks, the feedforward signal is often constructed as a filter acting on the setpoint, resulting in a feedforward controller with parameters describing this filtering operation. In this case, feedforward controller tuning is reduced to a parameter estimation problem given data of the system, either by estimating and inverting a system model \inputnewtext{\cite{DeRozario2017a}} or by directly estimating the inverse \cite{Blanken2020a}.

The key aspect determining the performance and task flexibility of the feedforward controller is the extent to which it can capture the relevant system dynamics \inputnewtext{\cite{Schoukens2019}}. For this purpose, parameterizations have become increasingly rich, ranging from polynomial and rational transfer functions for LTI systems \cite{Zou2004, Devasia1996}, nonlinear finite impulse response (NFIR) filters \cite{Bolderman2021}, and nonlinear autoregressive exogenous (NARX) and recurrent neural network (RNN) structures \cite{Sjoberg1995, Chen1990, Ljung2020}. While additional richness can increase performance, they also complicate estimation and may result in \inputnewtext{instability.}

Linear parameter-varying (LPV) parameterizations can capture a broad range of system dynamics and are promising to be deployed in feedforward control \cite{Bloemers2018}. LPV systems maintain linearity in signal relations, but the coefficients governing these signal relations are a function of time-varying scheduling variable $\rho$ that is assumed to be measurable online \cite{Toth2010}. LPV systems can embed nonlinear/ time-varying effects by proper selection of this scheduling variable \cite{Leith2000a}. \inputnewtext{Given such a signal}, the resulting linearity simplifies identification, stability analysis, and performance characterization.

In data-driven design of %parameter estimation for 
LPV feedforward control, it is desired to 1) employ a stochastic framework with generic noise models \cite{Toth2012} for consistency guarantees, 2) employ input-output (IO) representations for straightforward inversion, and, most importantly, 3) employ neural networks (NN) to address arbitrary (dynamic) coefficient dependency on $\rho$ and high-dimensional data.

In \cite{Bamieh2002, Zhao2012a}, the coefficients describing the LPV system are represented as a basis function expansion that is linear in the expansion coefficients, addressing 1) and 2). Yet, these suffer from the curse of dimensionality in high-dimensional input spaces \cite{Sjoberg1995}, may require more parameters \cite{Barron1994}%You: to describe the same dependence with a certain level of error
, and may be ill-conditioned, and thus fail to meet 3). A coefficient dependency parameterized by a neural network is considered in \cite{Previdi2003, Verhoek2022}, and by a Gaussian process in \cite{VanHaren2022a}, partially addressing 3). However, \cite{Previdi2003} considers an ARX noise model, considerably simplifying estimation, but often resulting in biased estimates, failing to meet 1). The method in \cite{VanHaren2022a} consider an LPV-NFIR feedforward parameterization, which also simplifies estimation, but limits the ability of the feedforward controller to compensate zero dynamics. In \cite{Verhoek2022}, general LPV system identification is carried out through a state-space (SS) parameterization where $\rho$ is synthesized using an encoder neural network from past IO data, effectively addressing 1) and 3). However, inversion of SS models %parameterizations 
is involved and \cite{Verhoek2022} is thus not suitable for 2). %especially for singular direct-feedthrough matrices,
%description with affine dependency to facilitate subsequent feedback controller synthesis. This 

Altough LPV feedforward control and related system identification approaches have addressed the separate aspects 1)-3), none of the existing methods meet all three requirements. Thus, the aim of this paper is to develop a tailored model structure for data-driven LPV feedforward design that meets all 3 requirements and exploits the mentioned developments.

The main contribution of this paper is the development of an NN-LPV-IO feedforward approach in which the coefficients depend on the scheduling variable $\rho$ through a neural network, and associated parameter estimation methods based on the prediction-error framework with generic noise models. The following subcontributions are distinguished.
\begin{enumerate}
	\item A local gradient-based optimization procedure for parameter estimation based on analytic gradient expressions. These expressions fully utilize the structure in the optimization problem, can be parallelized and are thus highly computationally efficient, as opposed to generic automatic differentiation frameworks \inputnewtext{(Section \ref{sec:local_optimiziation})}.
	\item A new iterative pseudolinear method for LPV output-error estimation, recovering the LTI Sanathanan-Koerner \cite{Sanathanan1963} method as a special case \inputnewtext{(Section \ref{sec:pseudolinear_regression})}. 
	\item \inputnewtext{A feedforward control simulation study illustrating that a NN parameterization of the coefficient functions results in superior performance compared to polynomial basis expansion coefficient functions (Section \ref{sec:simulation_example}).}
\end{enumerate}
%by reducing it to a static optimization as opposed to rolling out the filter response recursively over time

\subsubsection*{Setting \& Notation} Systems are discrete-time with sampling time $T_\mathrm{s}$. \inputnewtext{$\delta = T_\mathrm{s}^{-1} (1 - q^{-1})$ is the backward difference (delta) operator with $q$ the forward shift operator. Given a matrix $A \in \mathbb{R}^{m \times n}$, $A_{*,j}$ is its $j^\text{th}$ column and $\vec{A} = \begin{bmatrix} A_{*,1}^\T & \cdots & A_{*,n}^\T \end{bmatrix}^\T \in \mathbb{R}^{nm}$ its vector representation.}

%% file: Sections/Problem_formulation.tex
\section{Problem Formulation}
\label{sec:problem_formulation}
\begin{figure}[t]
\centering
\includegraphics{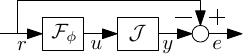}\vspace{-0mm}
\caption{Feedforward control configuration with system $\mathcal{J}$, output $y$, reference $r$, and error $e$. The input $u$ is parameterized as the output of a reference dependent filter $\mathcal{F}_{\phi}$.}\vspace{-5mm}
\label{fig:feedforward_setup}
\end{figure}
This section formalizes the considered feedforward control problem for LPV dynamic systems. Next, the solution is searched for in terms of a feedforward controller, parameterized as an LPV-IO difference equation with neural network coefficient functions. Lastly, given data obtained from the system, optimization of the defined feedforward controller is posed as an inverse system identification problem.

\subsection{Feedforward Control and Considered System Class}
The goal of feedforward control is to apply input $u(k) \in \mathbb{R}$ to a discrete-time system $\mathcal{J}$ such that its output $y(k) = \mathcal{J}(u(k)) \in \mathbb{R}$ is equal to some desired output $r(k) \in \mathbb{R}$, i.e.,
\begin{equation}
	e(k) = r(k) - y(k) = 0, \quad \forall k \in \mathbb{Z}_{> 0},
\end{equation}
where $e(k) \in \mathbb{R}$ is the error, see Fig.~\ref{fig:feedforward_setup}. $\mathcal{J}$ is considered as a deterministic LPV system given by the IO representation
\begin{equation}
    \sum_{i=0}^{N_a-1} a_i^t(\rho(k)) \delta^i y(k) = u(k) + \sum_{i=1}^{N_b-1} b_i^t(\rho(k)) \delta^i u(k), \label{eq:system}
\end{equation}
where $N_a,N_{b}\in\mathbb{Z}_{\geq 0}$ and, under the minimality conditions for \eqref{eq:system} \cite{Toth2010}, $\max(N_a,N_b)$ is the system order. $a_i^t,b_i^t$ are arbitrary coefficient functions with appropriate dimensions, dependent on the \emph{scheduling signal} $\rho(k) \in \mathbb{P} \subset \mathbb{R}^{N_\rho}$. This scheduling signal $\rho$ can be interpreted as describing the variation in operating conditions. By selecting it properly, \eqref{eq:system} can embed a large class of nonlinear systems \cite{Toth2010}. For example, $a_0^t(\rho) \!= \!k\rho^2$, $a_1^t \!=\! d$ and $a_2^t \!=\! m$ with $k,d,m \!\in \! \mathbb{R}_{>0}$ and $\rho = y$ represents a system with a cubic spring.

To ensure task flexibility of the feedforward methodology, $u$ is parameterized as the output of a feedforward controller operating on reference $r$ as specified next.
%parameterized as a filter acting on $r$
\subsection{Neural Network LPV-IO Feedforward Parameterization}
\begin{figure}[t]
\centering
\includegraphics[width=\linewidth]{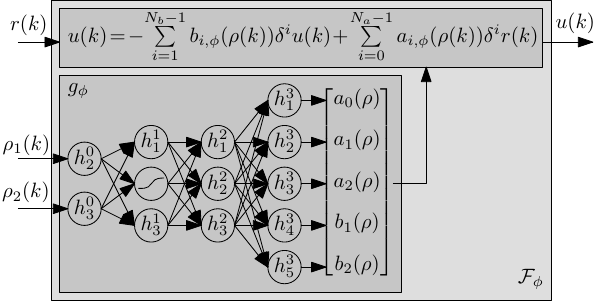} \vspace{-2mm}
\caption{Neural network LPV IO (NN-LPV-IO) feedforward parameterization, mapping desired outputs $r$ into feedforward signal $u$. Neural network $g_\phi$ transforms scheduling signal $\rho$ into parameter-varying coefficients $a_{i,\phi},b_{i,\phi}$ of the LPV difference equation. In this example, $g_\phi$ is drawn with dimensions $N_\rho = 2$, $N_a = 3$, $N_b = 3$ as a neural network with $L=2$ hidden layers, both of size $N_{l} = 3$, $l=1,2$. }
\label{fig:feedforward_parameterization} \vspace{-4mm}
\end{figure}

The considered feedforward controller $\mathcal{F}_\phi: r(k) \rightarrow u(k)$  with parameters $\phi$, visualized in  Fig. \ref{fig:feedforward_parameterization}, generates inputs $u(k)$ based on $r(k)$ according to the DT difference equation%
% Inkorten: hele DT DE weghalen, 'inputs' and 'referneces' weg, scheelt regel
\begin{equation}
\hspace{-2pt} u(k) + \sum_{i=1}^{N_b-1} b_{i,\phi}(\rho(k)) \delta^i u(k) = \sum_{i=0}^{N_a-1} a_{i,\phi}(\rho(k)) \delta^i r(k).
\label{eq:feedforward_parameterization}
\end{equation}
To learn systems $\mathcal{J}$ with arbitrary dependence of $a_i^t,b_i^t$ on $\rho$, the coefficient functions $a_{i,\phi}(\rho),\ b_{i,\phi}(\rho)$ are parameterized by a neural network $g_\phi$ acting on $\rho$ as defined next.
\begin{definition}
\label{def:scheduling_coefficients}
The coefficient functions are given by
\begin{equation}
\label{eq:NN_scheduling_coefficients}
\begin{bmatrix} a_{0,\phi}(\rho(k)) & \cdots & b_{N_b-1,\phi}(\rho(k)) \end{bmatrix}^\T = g_\phi(\rho(k)),
\end{equation}
with $g_\phi: \mathbb{R}^{N_\rho} \rightarrow \mathbb{R}^{N_a + N_b - 1}$ a neural network that corresponds to the mapping %generates coefficient functions according to%
%for input $x \in \mathbb{R}^{N_\rho}$
\begin{align}
&h^0(x) = x \nonumber \\
&\begin{aligned}
	z^l(x) &= W^{l-1} h^{l-1}(x) + c^{l-1} \\
	h^l(x) &= \kappa(z^l(x))
\end{aligned}
& \textrm{for  } l \in\{ 1, \ldots, L\} \label{eq:g_phi} \nonumber \\
&g_\phi(x) = W^{L} h^{L}(x) + c^{L} 
\end{align}
with hidden layer activation values $h^l \in \mathbb{R}^{N_l}$, bias vectors $c^l \in \mathbb{R}^{N_l+1}$, pre-activation values $z^l \in \mathbb{R}^{N_l}$ and weight matrices $W^l \in \mathbb{R}^{N_{l+1} \times N_l}$,  
with $N_0 = N_\rho$, $N_{L+1} = N_a + N_b - 1$, while $\kappa$ is an elementwise nonlinear activation function such as $\tanh$ or $\max(x,0)$. The parameter vector $\phi \in \mathbb{R}^{N_\phi}$ of $g_\phi$ is given by
\begin{equation}
\phi = \begin{bmatrix} \vec{W_0}^\T & c_0^\T & \ldots & \vec{W_L}^\T & c_{L}^\T \end{bmatrix}^\T .
\label{eq:phi}
\end{equation}
\end{definition}

\begin{remark}
Here $g_\phi$ is a fully connected multilayer perceptron, but\ifbool{long}{\textcolor{red}{, without loss of generality,}}{} it can be replaced by any \inputnewtext{feedforward} network, including user-defined input transformations \ifbool{long}{\textcolor{red}{acting on}}{}\ifbool{long}{\textcolor{blue}{of}}{of} $\rho$.
\end{remark}
\begin{remark}
    For a general representation of LPV systems with IO representations, it is necessary to also include a dependency of $a_{i,\phi},b_{i,\phi}$ on the time shifts of $\rho$, called dynamic dependency \cite{Toth2010}. This can easily be incorporated in \eqref{eq:feedforward_parameterization} by including also time-shifted $\rho$ as input to $g_\phi$. Here, only static dependency is considered for readability.
\end{remark}

Feedforward parameterization \eqref{eq:feedforward_parameterization} can represent a wide range of systems $\mathcal{J}$ under the common assumption of constant relative degree of $\mathcal{J}$ \cite{Bloemers2018}. More specifically, due to the universal function approximator property of neural networks \cite{Goodfellow2016}, they allow for representing any continuous scheduling dependence of $a_i^t, b_i^t$ on $\rho(k)$. Thus, given the correct order $N_a,N_b$ and enough parameters in $g_\phi$, \eqref{eq:feedforward_parameterization} can represent any system for which there exists an LPV-IO representation in the form of \eqref{eq:system}, including %systems in observable LPV SS form \cite{Toth2010} and 
a wide range of nonlinear systems.
% \begin{remark}
%     \textcolor{blue}{Since prior knowledge about the considered system is often available, a purely neural network approach might be unnecessary complex. Future work focuses on combining the neural network scheduling functions with a physical-model-based component to embed prior knowledge.}
% \end{remark}
%Moreover, any observable state-space (SS) LPV system has an equivalent IO representation \cite{Toth2010}, and can thus be represented by \eqref{eq:feedforward_parameterization}. The dependency on derivatives of $\rho$ introduced in the conversion from a SS to an IO representation can be addressed by including derivatives of $\rho$ in the input to $g_\phi$. Lastly, since LPV systems can embed nonlinear systems \cite{Toth2010}, \eqref{eq:feedforward_parameterization} can also capture a wide range of nonlinear systems under the assumption that the specific signal $\rho$ necessary for this embedding can be measured or estimated online.

% Dynamic scheduling dependency \cite{Toth2010}, i.e., dependency on derivatives of $\rho$, possibly introduced in the conversion from a state-space to an IO representation, can be addressed by augmenting $\mathbb{P}$ through including derivatives of $\rho$ in the input to $g_\phi$.
%Since any LPV system has an equivalent IO representation \cite{Toth2010}, possibly with the added complexity of dynamic scheduling, \eqref{eq:feedforward_parameterization} combined with \eqref{eq:g_phi} can capture any LPV system

\subsection{Inverse System Identification Problem}
\begin{figure}[t]
\centering
\includegraphics[]{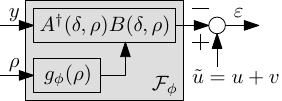}
\caption{Inverse system identification \inputnewtext{setup} to estimate parameters $\phi$ of the NN-LPV-IO feedforward controller \eqref{eq:feedforward_parameterization} by minimizing the $\ell_2$ loss of the generated output of $\mathcal{F}_\phi(y)$ with respect to the required input $u$, corrupted by white-noise $v$. }
\label{fig:inverse_system_identification_setup} \vspace{-3mm}
\end{figure}
To identify parameters $\phi$ of $\mathcal{F}_\phi$, an inverse system identification approach is adopted, \inputnewtext{see Fig. \ref{fig:inverse_system_identification_setup}}. More specifically, $\phi$ is optimized such that given a desired plant output $r$, $\mathcal{F}_\phi(r)$ generates the input $u$ required to obtain $r$. To this end, open-loop data $\mathcal{D} = \{ \tilde{u}(k), y(k), \rho(k)\}_{k=1}^N$ of $\mathcal{J}$ is considered to be available under the following assumptions.
\inputnewtext{
\begin{enumerate}[label=A\arabic*)]
    \item The measurement $\rho(k)$ is exact, i.e., it is not disturbed by noise.
    \item The measurement $y(k)$ is exact, and the measurement $\tilde{u}(k)$ of ${u}(k)$ is corrupted by an i.i.d. white noise $v(k)$, $\tilde{u}(k) = u(k) + v(k)$, with $v(k) \in \mathbb{R}$ and $\mathbb{E}(v^2) = \sigma_v^2$.
\end{enumerate}
}
Assumption A1 is commonly taken in the LPV literature and allows for a closed-form expression of the one-step-ahead predictor in the LPV case \cite{Toth2012}. Additionally, Assumption A2 creates an %NN-
LPV \emph{output-error} (OE) inverse identification setting in which the measurement of the input of the data-generating system is corrupted by noise, whereas the output can be exactly measured\footnote{Note that this setting is taken for the sake of simplicity, as if $y$ is perturbed by a noise process then the resulting noise structure in the inverse setting has an LPV Box-Jenkins form, where the input $y$ is correlated with the noise process. However, in the envisioned application domain of motion systems, signal-to-noise ratios are often high, such that this effect is limited.}.
Under these assumptions, the optimal one-step-ahead predictor $\uhat$ is given by \cite{Toth2012}
\begin{equation}
	\uhat = B^\dag(\delta,\rho(k)) A(\delta,\rho(k)) y(k) = \mathcal{F}_\phi(y(k)),
	\label{eq:one_step_ahead_prediction}
\end{equation}
with 
\begin{equation}
\begin{aligned}
	A(\delta,\rho(k)) &= \sum\nolimits_{i=0}^{N_a-1} a_{i,\phi}(\rho(k)) \delta^i \\
	B(\delta,\rho(k)) &= 1 + \sum\nolimits_{i=1}^{N_b-1} b_{i,\phi}(\rho(k)) \delta^i,
	 \label{eq:B_phi}
\end{aligned}
\end{equation}
with $B^\dag(\delta,\rho)$ being the inverse of the monic $B(\delta,\rho)$, i.e, $B^\dag(\delta,\rho) B(\delta,\rho) u = u$, such that $B^\dag(\delta, \rho(k)) A(\delta,\rho(k))$ represents the filtering operation \eqref{eq:feedforward_parameterization}, see \cite{Toth2012} for details of this inversion operation.
%and $B^\dag(\delta,\rho)$ is the convergent adjoint of $B(\delta,\rho)$, i.e., $B^\dag(\delta,\rho) B(\delta,\rho) = 1$.  
The parameters $\phi$ are optimized in terms of minimizing the $\ell_2$-loss of the one-step-ahead prediction %$\hat{u}$ 
of $u$ by $\mathcal{F}_\phi$ given $y$, i.e., according to $\phi^* = \arg \min V(\phi)$ with \vspace{-2mm}
\begin{equation}
V(\phi) = \sum_{k=1}^N \varepsilon^2(k) = \sum_{k=1}^N \left(\tilde{u}(k) - \uhat \right)^2.
\label{eq:V_OE} \vspace{0.1mm}
\end{equation}

\begin{remark}
The inverse system identification setup and subsequent results can easily be extended to settings in which $v$ is not white by introducing a parameterized noise model of $v$ as in \cite{Toth2012,Zhao2012a} to capture the noise spectrum. % and modifying \eqref{eq:one_step_ahead_prediction} as in \cite{Zhao2012a}. 
\end{remark}
The goal of this paper is to develop algorithms for optimization of \eqref{eq:V_OE}, i.e., for determining $\phi^\ast$ efficiently. To this extent, the structure of $V(\phi)$ and its analytic gradient are fully utilized and combined with a second order solver. In contrast, standard automatic differentiation frameworks do not exploit the structure of this cost function.
% Subsequently, the optimized feedforward controller can be used to improve tracking performance.

%% file: Sections/Local_optimization.tex
\section{Local Optimization}
\label{sec:local_optimiziation}
The first strategy for optimization of \eqref{eq:V_OE} is through local optimization using gradient-based algorithms. Consequently, first the gradient of the cost function $V(\phi)$ and the Jacobian of $\varepsilon(k)$ with respect to $\phi$ are derived. Second, it is shown how to use those in local optimization using the Levenberg-Marquardt algorithm. These steps fully define any gradient-based optimization, constituting Contribution 1.

\subsection{Gradient and Jacobian of $V(\phi)$} 
The gradient $\gradphij V(\phi) \in \mathbb{R}$ of \eqref{eq:V_OE} with respect to a single parameter $\phi_j \in \mathbb{R}$ of \eqref{eq:phi} is given by
\begin{equation}
	\gradphij V(\phi) = 2 \sum_{k=1}^N J_{\phi_j}(k) \varepsilon(k),
	\label{eq:grad_phi_V_OE}
\end{equation}
%\sum_{k=1}^N \frac{d \varepsilon(k)}{d \phi_j} \frac{d \varepsilon^2(k)}{d \varepsilon(k)}
where $J_{\phi_j}(k) = \frac{d \varepsilon(k)}{d \phi_j} \in \mathbb{R}$ is the gradient of $\varepsilon$ at time step $k$ with respect to $\phi_j$. Deriving an expression for the neural network parameter gradients $\frac{d a_i(\rho(k))}{d \phi_j}$, $\frac{d b_i(\rho(k))}{d \phi_j}$ enables calculating this gradient\inputnewtext{, as} given by the following two lemmas.

\begin{lemma} %sensitivities
\label{lem:backprop}
Given $g_\phi(x)$ in \eqref{eq:g_phi}, define $\Delta^l = \frac{d g_\phi(x)}{d z^l(x)}$ with $l=1,\ldots,L$. Then the output layer sensitivity $\Delta^L = \frac{d g_\phi(x)}{d z^L(x)} = I_{N_a + N_b -1}$, and subsequent layers satisfy 
\begin{equation}
	\Delta^{l-1} = \frac{d g_\phi(x)}{dz^l(x)} \frac{d z^l(x)}{d z^{l-1}(x)} = \Delta^l \frac{d \kappa(z^l(x))}{d z^l(x)} W^{l-1},
	\label{eq:backprop}
\end{equation}
with $\frac{d \kappa(z^l(x))}{d z^l(x)}$ the diagonal matrix of derivatives of $\kappa$, i.e.,
\begin{equation*}
	\frac{d \kappa(z^l(x))}{d z^l(x)} = \diag (\begin{bmatrix} \frac{d \kappa(\eta)}{d\eta}\vert_{\eta = z^l_1(x)} & \dots & \frac{d \kappa(\eta)}{d\eta}\vert_{\eta = z^l_{N_l}(x)} \end{bmatrix}).
\end{equation*}
\end{lemma} 

Lemma \ref{lem:backprop} defines the  hidden layer sensitivities $\Delta^l = \frac{d g_\phi(x)}{d z^l(x)}$ of the neural network as a recurrence through the layers, which % $l$. 
%It 
can be seen as %recognized as the 
backpropagation %{algorithm} 
applied to all $N_a + N_b - 1$ outputs of $g_\phi$ simultaneously. Then, the parameter 
gradients $\frac{d g_\phi(x)}{d\phi_j}$ %at each layer 
%then 
can be computed from $\Delta^l$.
%follow from $\Delta^l$, as formalized next.%
%
\begin{lemma}
Given $\Delta^l$ as in Lemma \ref{lem:backprop} and $e_i \in \mathbb{R}^{N_a + N_b - 1}$ as the unit vector with its $i^\text{th}$ entry equal to 1. The parameter gradients $\frac{d a_{i,\phi}(x)}{d W^l_{m,n}}, \frac{d a_{i,\phi}(x)}{d c^l_m}$ are given by
%\frac{d g_\phi(x)}{d z^l(x)} \frac{d z^l}{d W^l_{n,m}} = 
\begin{equation} \label{eq:bp:full}
	\begin{aligned}
		\frac{d a_{i,\phi}(x)}{d W^l_{m,n}} &= e_i^\T \frac{d g_\phi(x)}{d W^l_{m,n}} = e_i^\T \frac{d g_\phi(x)}{d z^{l+1}(x)} \frac{d z^{l+1}(x)}{d W^l_{m,n}} \\
		&= e_i^\T \Delta^{l+1} \frac{dW^{l} h^{l} + c^{l}}{dW^l_{m,n}} = \Delta^l_{i,m} h^{l-1}_n, \\
		\frac{d a_{i,\phi}(x)}{d c^l_m} &= e_{i}^\T \Delta^{l+1} \frac{dW^{l} h^{l} + c^{l}}{d c^l_m} = \Delta^l_{i,m},
	\end{aligned}
\end{equation}
and similarly for $\frac{d b_{i,\phi}(x)}{d W^l_{m,n}}, \frac{d b_{i,\phi}(x)}{d c^l_m}$ with $e_{i+N_a}$. 
\label{lem:NN_parameter_gradients}
\end{lemma}%
%the residual Jacobian
With the above lemmas,  $J_{\phi_j}(k)$ is computed by an LPV filtering operation as defined next.
\begin{theorem}
	Given $\frac{d a_{i,\phi}}{d \phi}, \frac{d b_{i,\phi}}{d \phi}$ in Lemma \ref{lem:NN_parameter_gradients}, $J_{\phi_j}(k)= \frac{d \varepsilon(k)}{d \phi_j}$ is given by
\begin{align}
	J_{\phi_j}(k) &= - \Big( B^\dagger(\delta,\rho(k)) \sum_{i=0}^{N_a-1} \frac{d a_{i,\phi}(\rho(k))}{d\phi_j} \delta^i 	\label{eq:residual_parameter_gradients}
\\
	- \big( B^\dagger&(\delta,\rho(k))\big)^2  \sum_{i=1}^{N_b-1} \frac{d b_{i,\phi}(\rho(k))}{d\phi_j} \delta^i A(\delta,\rho(k)) \Big) y(k), \nonumber
\end{align}
where $\left(B^\dagger(\delta,\rho(k))\right)^2$ denotes filtering with  $B^\dagger(\delta,\rho(k))$ twice.
\label{th:residual_parameter_gradients}
\end{theorem}

\begin{remark}
Jacobian computation \eqref{eq:residual_parameter_gradients} can be parallelized for all $\phi_j$ and, except for the inverse operation $B^\dag$, for all times $k$. The vectorized expressions are omitted for brevity, but can be found in the software implementation \cite{gitlab_lpv_io_identification}.
\end{remark}
Theorem \ref{th:residual_parameter_gradients} provides an expression for the Jacobian $J_{\phi_j}(k)$ that is part of the gradient of $V(\phi)$, see \eqref{eq:grad_phi_V_OE}, completing the full gradient derivation. These analytic expressions are more computationally efficient than automatic differentiation, enabling fast optimization of $V(\phi)$.

\subsection{Optimization Using Levenberg-Marquardt}
The gradient expressions \eqref{eq:grad_phi_V_OE}-\eqref{eq:residual_parameter_gradients} can be used with the \emph{Levenberg-Marquardt} (LM) algorithm  to optimize $V(\phi)$. LM is a method specifically tailored for solving problems such as $\arg \min_\phi V(\phi)$ by exploiting its least-squares structure. Thus, it can reach quadratic convergence rate through a Gauss-Newton approximation of the Hessian. LM generally outperforms first order schemes such as accelerated/conjugate gradient descent in terms of floating point operations required for convergence \cite{Hagan1994}. 

More specifically, define the parameter Jacobian \textit{matrix} $J(\phi) \in \mathbb{R}^{N \times N_a + N_b - 1}$ such that $J_{k,j}(\phi) = J_{\phi_j}(k)$ with $J_{\phi_j}(k)$ in \eqref{eq:residual_parameter_gradients}. Then, the LM algorithm proceeds as follows.%
\vspace{3pt}
\hrule
\vspace{-2pt}
\begin{procedure}[LM for NN-LPV-IO OE optimization] \hfill
	\vspace{3pt} \hrule \vspace{2pt}
	Given model structure \eqref{eq:feedforward_parameterization} with parameters $\phi$, and dataset $\mathcal{D}$, set $q=1$ and initialize $\phi^{0},\lambda>0,\mu>0$. Then, iterate:
	\begin{enumerate}[label=(\arabic*)]
        \item Calculate $\varepsilon_1^N$ according to \eqref{eq:one_step_ahead_prediction}-\eqref{eq:V_OE}.
		\item Calculate $J(\phi_{q-1})$ according to \eqref{eq:backprop}-\eqref{eq:residual_parameter_gradients}.
		\item Determine $\phi_{q}$ as 
		\begin{equation}
				\phi_q = \phi_{q-1} - \alpha_{q-1} H(\phi_{q-1})^{-1} g(\phi_{q-1}),
				\label{eq:local_update}
		\end{equation}
		with $\alpha_{q-1} = 1$, $H(\phi_{q-1}) = J(\phi_{q-1})^\T J(\phi_{q-1}) + \lambda I$, and $g(\phi_{q-1}) = -J(\phi_{q-1})^\T \varepsilon_1^N$.
		\item If $V(\phi_q) > V(\phi_{q-1})$, set $\lambda = \mu \lambda$ and continue to (5). Else, set $\lambda = \mu^{-1} \lambda$ and go to (3). Typically $\mu=10$.
		\item Set $q = q + 1$ and go back to (1) until convergence, e.g., until $\abs{\phi_{q} - \phi_{q-1}} < \eta$ with tolerance $\eta$ , or $q = q_{max}$.
	\end{enumerate}
	\label{proc:LM_iterations}
	\vspace{0pt} 	\hrule 	\vspace{-2pt}
\end{procedure}

Update \eqref{eq:local_update} can be recognized as a damped Gauss-Newton step, interpolating between gradient descent and an approximate second order method, and is successfully applied to LPV-IO estimation with polynomial scheduling in \cite{Zhao2012a}. 
\begin{remark}
	Other gradient-based schemes can be also derived based on \eqref{eq:local_update} with different choices for $H(\phi^n)$ and $g(\phi^n)$, e.g., $H(\phi^n) = I$ corresponds to gradient descent. %\textcolor{blue}{For example, $H(\phi^n) = I$ corresponds to gradient descent. Quasi-newton methods like BFGS iteratively update $H(\phi^n)$ based on $g(\phi^n)$, and Newton methods use the exact Hessian.}
\end{remark}

%% file: Sections/Pseudolinear_regression.tex
\section{Pseudolinear Regression}
\label{sec:pseudolinear_regression}
This section introduces a different algorithm for optimization of \eqref{eq:V_OE} of which the LTI counterpart is often used to generate initial estimates for the local optimization of Section \ref{sec:local_optimiziation}\deleteoldtext{\cite{Pintelon2012_Ch98}}, constituting Contribution 2.\deleteoldtext{The key idea of this algorithm is to keep components of \eqref{eq:V_OE} fixed at their value in the previous iteration, such that at each iteration $B^\dag(\delta,\rho(k))$ enters the optimization in a static as opposed to a recurring manner. Consequently, gradient computation is considerably simplified with respect to Theorem \ref{th:residual_parameter_gradients}, such that parameter updates can be computed faster.} For compactness, the dependence of $y,u,\rho$ on $k$ is omitted in this section.

\subsection{Sanathanan-Koerner Iterations}
\inputnewtext{The key idea of Sanathanan-Koerner (SK) iterations \cite{Sanathanan1963} is that the filtering operation $B^\dag(\delta,\rho)$ is treated as an a priori unknown weighting and updated iteratively. Specifically, $V(\phi)$ can be rewritten as}
\begin{equation}
    V(\phi) = \sum\nolimits_{k=1}^N \Big( B^\dag(\delta,\rho) \left(B(\delta,\rho) \tilde{u} - A(\delta,\rho) y \right) \Big)^2.
    \label{eq:V_rewritten}
\end{equation}
Now define iteration counter $q$ and denote by $B_{q-1}^\dag(\delta,\rho)$ the filter corresponding to variables $\phi_{q-1}$, i.e., $B^\dag(\delta,\rho)$ in iteration $q-1$. Next, $B^\dag(\delta,\rho)$ is fixed at $B_{q-1}^\dag(\delta,\rho)$ to obtain
\begin{equation}
	\hspace{-.5em} V^q_\mathrm{SK}(\phi) \! = \! \sum_{k=1}^N \left(\! B^\dag_{q-1}(\delta,\rho) \left(B(\delta,\rho) \tilde{u} - A(\delta,\rho(k)) y \right) \!\right)^2 \hspace{-.5em},
	\label{eq:V_SK}
\end{equation}
such that $V_\mathrm{SK}^{q}(\phi)$ is equal to $V(\phi)$ if $B_{q-1}(\delta,\rho) = B(\delta,\rho)$. 
$V_\mathrm{SK}^{q}$ can be recognized as an ARX criterion weighted by $B^\dag_{q-1}(\delta,\rho)$, where $B^\dag_{q-1}(\delta,\rho)$ is updated over the iterations.

At every iteration $q$, \eqref{eq:V_SK} is minimized over $\phi$. Even though nonlinear optimization is still required for this, this is a cheap operation compared to minimizing the full $\ell_2$ loss \eqref{eq:V_OE}, see below. Analogously to \eqref{eq:residual_parameter_gradients}, the gradient of \eqref{eq:V_SK} is given by
\begin{multline}
    \hspace{-1em}\frac{d}{d\phi_j} V_\mathrm{SK}^q(\phi) = 2 \sum_{k=1}^N \left(\! B^\dag_{q-1}(\delta,\rho) \left(B(\delta,\rho) \tilde{u} - A(\delta,\rho(k)) y \right) \!\right) \\
    \cdotp B^\dag_{q-1}(\delta,\rho) \big(\! \sum_{i=1}^{N_b-1} \frac{db_{i,\phi}(\rho)}{d \phi_j} \delta^i \tilde{u} - \sum_{i=0}^{N_a-1} \frac{da_{i,\phi}(\rho)}{d \phi_j} \delta^i y \big),\label{eq:SK_gradient}
\end{multline}
in which the first line represents the SK residuals $\varepsilon^q_\mathrm{SK}(k)$, and the second line the gradient of $\varepsilon^q_\mathrm{SK}(k)$ with respect to $\phi_j$. Iterative criterion \eqref{eq:V_SK} and gradient \eqref{eq:SK_gradient} give rise to the LPV-SK-algorithm for optimizing $V(\phi)$, as formalized next.

\vspace{3pt}
\hrule
\vspace{-2pt}
\begin{procedure}[SK for NN-LPV-IO OE optimization] \hfill
	\vspace{3pt} \hrule \vspace{2pt}
	Given model structure \eqref{eq:feedforward_parameterization} with parameters $\phi$, and dataset $\mathcal{D}$, set $q=1$, $B_{0}(\delta,\rho) = 1$ and initialize $\phi^{0}$. 
	Then, iterate:
	\begin{enumerate}[label=(\arabic*)]
		\item Given $\phi_{q-1}$ determine $\phi_{q}$  as
		\begin{equation}
			\phi_q = \arg \min_\phi V_\mathrm{SK}^q(\phi) \label{eq:V_SK_explicit},
		\end{equation}
        using \eqref{eq:V_SK} and \eqref{eq:SK_gradient}.
		\item Set $q = q + 1$ and go back to (1) until convergence.
	\end{enumerate}
	\label{proc:SK_iterations}
	\vspace{0pt} 	\hrule 	\vspace{-2pt}
\end{procedure}
\begin{remark}
If coefficients $a_i(\rho),b_i(\rho)$ are parameterized linearly in the parameters $\phi$, \eqref{eq:V_SK_explicit} reduces to a weighted linear least-squares problem as in the LTI case, see \cite{gitlab_lpv_io_identification}.
\end{remark}

\inputnewtext{Algorithm \ref{proc:SK_iterations} is potentially more computationally efficient than Algorithm \ref{proc:LM_iterations}. Parameters $\phi$ in $B(\delta,\rho)$ enter \eqref{eq:V_SK} as $B(\delta,\rho)u$ as opposed to as $B^\dag(\delta,\rho)A(\delta,\rho)y$ in \eqref{eq:V_OE}, i.e., in a non-recurring manner. Consequently, the gradient computation is computationally cheaper: \eqref{eq:SK_gradient} contains only a single inverse LPV filtering operation ($B^\dag_{q-1}$), whereas \eqref{eq:grad_phi_V_OE} contains at least two, thus obtaining cheaper parameter updates at the cost of having to iterate over $q$. }

\inputnewtext{The SK algorithm for LPV-OE identification has not been considered in literature before. Already in the LTI case, convergence of Algorithm \ref{proc:SK_iterations} is not guaranteed, but good convergence is observed in a range of applications especially for high signal-to-noise ratios. The convergence for the LPV case is investigated in simulation in Section \ref{sec:simulation_example}.}

%% file: Sections/Simulation_example.tex
\section{Simulation Example}
\label{sec:simulation_example}
In this section, performance of the proposed feedforward learning approach
%\eqref{eq:feedforward_parameterization} 
is investigated on a simulation example of an %validated through a simulated 
LPV system. It is shown that \eqref{eq:feedforward_parameterization} can learn from IO data both the required input as well as the true scheduling dependencies. %map trained on data from a simulated LPV system. 
Additionally, it is compared to standard methods utilizing a polynomial parameterization of the scheduling dependence, where the proposed NN-LPV-IO method shows superior performance%parameterization
, constituting Contribution 3.

\subsection{Example System}
\begin{figure}[ht]
\centering
\includegraphics[]{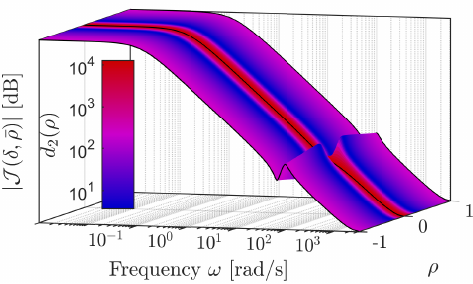} \vspace{-2mm}
\caption{Bode plot of frozen dynamics of $\mathcal{J}(\delta,\rho)$, displaying a resonance with scheduling-dependent damping.}
\label{fig:LPV_Bode} \vspace{-2mm}
\end{figure}

As an example of \eqref{eq:system}, consider a true LPV system $\mathcal{J}: u(k) \rightarrow y(k)$\inputnewtext{, see Fig. \ref{fig:LPV_Bode},} that is governed by
\begin{gather}
    k_1 k_2 y(k) + \Big(\!d_1 k_2 \!+\! k_1 d_2(\rho)\!\Big) \delta y(k) + \Big(\!k_1 m_2 \! + \! k_2 (m_1 \!+\! m_2) \nonumber \\
    \!+\! d_1 d_2(\rho)\!\Big) \delta^2 y(k) + \Big(\!d_1 m_2\!+\! (m_1\!+\!m_2) d_2(\rho)\!\Big) \delta^3 y(k)     \label{eq:example_system}
 \\
    + m_1 m_2 \delta^4 y(k)  
    = k_2 u(k) +  d_2(\rho) \delta u(k) + m_2 \delta^2 u(k), \nonumber
\end{gather}
with parameters $m_1=10,m_2 = 11$ kg, $d_1=5$ Ns/m, $k_1=0.3,k_2=5\cdot 10^4$ N/m. The system is subject to damping that depends nonlinearly on the scheduling $\rho$ as
\begin{equation}
	d_2(\rho) = 10^3 \exp{(-10^2 \rho^2)} + 10^2 \rho^2 + 10^{-4}.
\end{equation}
\deleteoldtext{This varying damping manifests as a resonance with $\rho$-dependent amplification as shown in Fig. \ref{fig:LPV_Bode}. For fixed $\rho$, \eqref{eq:example_system} can be recognized as the transfer function of a two-mass-damper-spring system with $d_2(\rho)$ describing the varying damping of the flexible mode, representing a situation with very high damping around $\rho = 0$.}

For this system, a dataset $\mathcal{D}$ of $N=6400$ samples is generated in which $y$ is a fourth order reference, while $u$ corresponding to this $y$ is obtained by simulating \eqref{eq:example_system} in reverse direction. Then, $u$ is perturbed by white measurement noise $v$ with $\mathbb{E}(v^2) = 0.01^2 \textrm{var}(u)$. Additionally, the scheduling trajectory is taken as $\rho(k) = 2\frac{k-1}{N-1} -1 \in [-1,1]$. 

\subsection{Comparison to polynomial coefficient functions}
$\mathcal{F}_\phi$ in \eqref{eq:feedforward_parameterization} is chosen with $N_a = 5$, $N_b = 3$ and a neural network according to Definition \ref{def:scheduling_coefficients} with $L=2$, $N_0=1,N_1=5,N_2=5,N_3=N_a + N_b - 1$, $\kappa=\tanh$, i.e., 2 hidden layers with 5 nodes and a linear output layer with $N_a + N_b - 1 = 7$ nodes, totaling $N_\phi = 82$ parameters. This parameterization has $N_a,N_b$ equal to the true order of \eqref{eq:example_system}, such that $\mathcal{F}_\phi$ can capture \eqref{eq:example_system} up to the approximation capabilities of $g_\phi$.

$\mathcal{F}_\phi$ is optimized over $\mathcal{D}$ using the gradient-based optimization of Section \ref{sec:local_optimiziation}. Initial parameters are obtained using an ARX criterion and lowpass filtered data. Additionally, $\mathcal{F}_\phi$ is compared to a feedforward parameterization $\mathcal{G}_\theta$ based on a linear combination of polynomial basis functions to capture the scheduling dependencies, i.e., $a_i(\rho) = \sum_{j=0}^{N_{a_i}} \alpha_i^j a_i^j(\rho)$, \inputnewtext{$b_i(\rho) = \sum_{j=0}^{N_{b_i}} \beta_i^j b_i^j(\rho)$}, resulting in 91 parameters \cite{Zhao2012a}.
\begin{figure}[ht]
\centering
\includegraphics[]{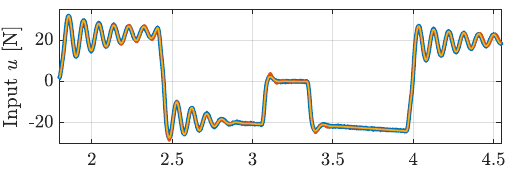}
\includegraphics[]{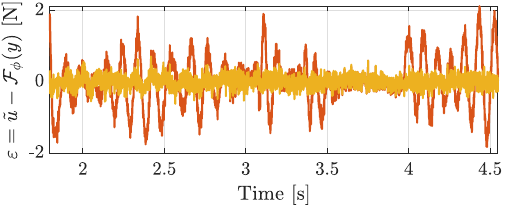} \vspace{-6mm}
\caption{The NN-LPV-IO feedforward model \eqref{eq:feedforward_parameterization} optimized with Algorithm \ref{proc:LM_iterations} is able to generate $\mathcal{F}_\phi(y)$ (\protect \drawlinelegend{morange}) that matches the measured $\tilde{u}(k)$ (\protect \drawlinelegend{mblue}) up to the noise level. In contrast, an LPV-IO feedforward parameterization based on polynomial basis function expansion (\protect \drawlinelegend{mred}) is not rich enough to capture all dynamics, resulting in coloured residuals. The input $u(k)$ clearly reflects LPV characteristics of the $\mathcal{J}(\delta,\rho)$, i.a., through the absence of oscillations for $\rho \approx 0$ around $t \in [3,3.5]$ by increased damping $d_2(\rho)$.}
\label{fig:input_and_residuals} \vspace{-2mm}
\end{figure}
\begin{figure}[ht]
\centering
\includegraphics[scale=1]{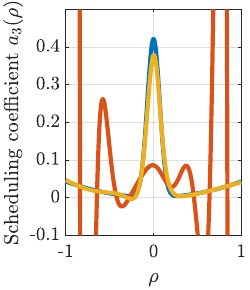}
\includegraphics[scale=1]{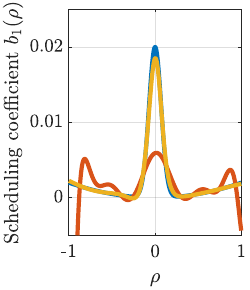} \vspace{-4mm}
\caption{The NN-LPV-IO model (\protect \drawlinelegend{morange}) is able to accurately learn the true coefficient functions (\protect \drawlinelegend{mblue}) through a neural network parameterization of the coefficient functions as in \eqref{eq:NN_scheduling_coefficients}. In contrast, a polynomial basis function expansion with the same number of coefficients (\protect \drawlinelegend{mred}) cannot capture the true coefficients \inputnewtext{due to the non-polynomial dependency on $\rho$.}}
\label{fig:scheduling_coefficients} \vspace{-2mm}
\end{figure}
\inputnewtext{The generated and measured $u$ are visualized in Fig. \ref{fig:input_and_residuals}. It indicates that $\mathcal{F}_\phi$ can accurately represent the true system through the use of a neural network, obtaining residuals $\varepsilon$ that are almost white with costs $N^{-1}V(\phi) = 0.028$ close to noise variance $\mathbb{E}(v^2) = 0.026$. In contrast, $\mathcal{G}_\theta$ cannot capture all dynamics due to the limitations of the polynomial representation, obtaining colored residuals with costs $N^{-1}V(\theta) = 0.486 \gg \mathbb{E}(v^2)$. For a validation dataset containing similar but different data with $\mathbb{E}(v^2) = 0.029$, $\mathcal{F}_\phi$ obtains $N^{-1} V(\phi) = 0.044$, compared to $N^{-1} V(\theta) = 1.13$ for $\mathcal{G}_\theta$.
Additionally, Fig. \ref{fig:scheduling_coefficients} shows the estimated $a_3(\rho)$, $b_1(\rho)$ with true coefficient functions $a_3^t(\rho) = \big(d_1 m_2 \! + \! (m_1\!+\!m_2) d_2(\rho)\big) k_2^{-1}$ and $b_1^t(\rho) = d_2(\rho) k_2^{-1}$. It is observed that $\mathcal{F}_\phi$ can effectively capture $a_3^t(\rho),b_1^t(\rho)$, whereas the polynomial parameterization fails to do so.}

% Test and validation error for neural network and polynomial coefficient functions. 
% \begin{table}[ht]
% \centering
% \caption{Error $\frac{1}{N} V(\phi)$ for $\mathcal{F}_\phi$ ($\mathcal{G}_\theta$) with coefficient functions $a_i(\rho),b_i(\rho)$ given by a neural network (polynomial).}
% \label{tab:training_losses}
% \begin{tabular}{@{}llll@{}}
% \toprule
%  & $\mathcal{F}_\phi$ & $\mathcal{G}_\theta$ & Noise variance $\mathbb{E}(v^2)$ \\ \midrule
% Training                       & 0.028 & 0.486      & 0.026                            \\
% Validation                        & 0.044 & 1.13       & 0.029                            \\ \bottomrule
% \end{tabular}
% \end{table}

\begin{figure}
\centering
\includegraphics{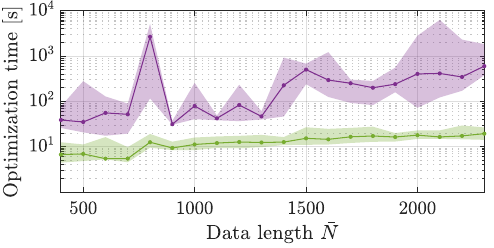} \vspace{-2mm}
\caption{\inputnewtext{Median (\protect\drawlinelegend{black}) and interval \protect\drawrectanglelegend{black}{30} of computation times of Algorithm \ref{proc:LM_iterations} (\protect\drawrectanglelegend{mgreen}{40}) and a vanilla implementation in Pytorch using ADAM (\protect\drawrectanglelegend{mpurple}{40}) for different data lengths $\bar{N}$ and $7$ random initializations. By incorporating analytic gradients and the problem structure, Algorithm \ref{proc:LM_iterations} is about 10 times faster.}}
\label{fig:training_times} \vspace{-2mm}
\end{figure}

\deleteoldtext{In general, this simulation example is representative regarding the flexibility of \eqref{eq:feedforward_parameterization} in learning arbitrary coefficient functions, significantly improving performance over basis function expansion methods. However, like any non-convex optimization method, it is sensitive to local minima, and stability of the learned filter $\mathcal{F}_\phi(y)$ is not guaranteed. }%, possibly resulting in unbounded \TR{responses of} $\mathcal{F}_\phi(y)$.
\inputnewtext{
\subsection{Comparison of Optimization Methods}
This section compares the computation times of the developed Algorithms \ref{proc:LM_iterations} and \ref{proc:SK_iterations}, and an implementation in Pytorch using automatic differentiation and ADAM. Specifically, dataset $\mathcal{D}$ is truncated at different lengths $\bar{N}_i < N$, and $\mathcal{F}_\phi$ is optimized according to $V(\phi)$ over this truncated data until $V(\phi)$ reaches threshold $\bar{N}_i^{-1} V(\phi) \leq 0.3^2 $. All experiments were carried out on a HP Z-book G5 using a Intel Core i7-8750H CPU. Fig. \ref{fig:training_times} shows the computation time for each method. The following observations can be made. 
\begin{itemize}
    \item Algorithm \ref{proc:SK_iterations} does not converge for the considered example for any data length, and is thus not shown. Although it converges sometimes for the polynomial coefficient functions \cite{gitlab_lpv_io_identification}, it is too brittle for the neural network coefficient function case.
    \item By including analytic gradients and (approximate) curvature of $V(\phi)$, Algorithm \ref{proc:LM_iterations} is about 10 times faster than automatic differentation with ADAM.
    \item For $N=6400$, Algorithm \ref{proc:SK_iterations} takes about 350 s, whereas optimization using automatic differentiation is too time consuming using an %with just an 
    average laptop.
\end{itemize}
}
%All routines have been provided with the same initial variables, namely the result of an ARX estimation. 

%% file: Sections/Conclusion.tex
\section{Conclusion}
\label{sec:conclusion}
This paper has introduced a feedforward controller structure %parameterization 
based on LPV input-output representations with a neural network describing the dependency of the the controller parameters %coefficients $a_i,b_i$ 
on the scheduling variable. %$\rho$}. T
The inclusion of LPV autoregressive dynamics in the feedforward controller directly allows for learning and compensating a wide class of systems, including those with shifting anti-resonances, in contrast to most methods in literature that only tackle shifting resonances through an LPV-NFIR feedforward parameterization. A computationally efficient method for learning the coefficient functions that utilizes the structure of the optimization problem is developed. Based on a simulation study, %The approach is validated in simulation, illustrating the 
increased performance and computational superiority of the developed NN-LPV-IO feedforward approach is demonstrated. \vspace{-3mm} %parameterization.